\documentstyle[12pt]{article}

\begin{document}
\title{\large \bf On the scale-invariant distribution of the diffusion
coefficie
nt for classical particles diffusing in disordered media.  }
\author{\normalsize Yan-Chr Tsai and Yonathan Shapir \\
        \normalsize Department of Physics and Astronomy\\
        \normalsize University of Rochester\\
        \normalsize Rochester, N. Y. 14627, U.S.A.}

\maketitle
\large
\begin{center} Abstract \end{center}
\small

 The scaling form of the whole distribution $P(D)$ of the random diffusion
 coefficient $D(\vec{x})$ in a model of classically diffusing particles is
investigated. The renormalization group approach above the lower
 critical dimension $d=0$ is applied to the distribution $P(D)$ using the
n-replica approach. In the annealed approximation $(n=1)$, the inverse gaussian
 distribution is found to be the stable one under rescaling. This
identification
 is made based on symmetry arguments and subtle relations between this model
 and that of fluctuating interfaces studied by Wallace and Zia.
 The renormalization-group flow for the ratios between subsequent cumulants
 shows a regime of pure diffusion for small disorder,  where
  $P(D) \rightarrow \delta (D-\bar{D}) $,  and a regime of strong disorder in
which the cumulants grow infinitely large and the diffusion process is ill
 defined. The boundary between these two regimes is associated with an
 unstable fixed-point and a subdiffusive behavior
 $<\vec{x} ^2 > \sim t^{1-d/2}$. For the quenched
($n \rightarrow 0$) case  we find that unphysical operators are generated
 raising doubts on the renormalizability of this model. Implications to other
 random systems near their lower critical dimension are discussed.
\normalsize
\hskip 1.0truein

\newpage

\noindent {\bf \large I. Introduction.}

  The asymptotic behavior of classical particles diffusing in a disordered
 media has been the focus of many recent investigations [1]. The models which
 have attracted most attention are formulated in terms of random transition
 probabilities or local random
  quenched forces. In most of these problems $d=2$ is a special critical
 dimension and most attention has been foucused on the anomalous diffusion
 which occurs usually for $ d<2$ [2-7]. The anomalous diffusion is manifested
 by the long-time behavior of the form:
\begin{equation}
\langle  \vec{x}^2(t)\rangle \sim t^{2\mu},\label{eq:asym}
\end{equation}
 with $2\mu <1 (2\mu >1)$, the behavior is said to be sub-(super-) diffusive.
   Attention has been devoted also to a simpler model in which the
 diffusion constant $ D(\vec x)$ is a local quenched random variable and the
 diffusion equation for the local density
$\rho (\vec x,t)$ is:
\begin{equation}
\frac{\partial\rho(\vec{x},t)}{\partial
t}=\vec{\nabla}\cdot[D(\vec{x})\vec{\nabla} \rho (\vec{x},t)]. \label{eq:diff}
\end{equation}

This equation describes as well the continuum limit of a random resistor
 network in which $D(\vec x)$ is the local conductance. In this context one
 can define an Hamiltionian ${\cal E}=\frac{1}{2}\int d^d x D(\vec
 x)(\vec{\nabla} V(\vec x))^2 $, where $V(\vec x)$ is the local voltage. The physical
 problem in which the $D(\vec x)$ are themselves fluctuating dynamical
 variables required averaging the partition function (annealed average),
while if the $D(\vec x)$ are frozen, it is the free-engery which should
be averaged (quenched average).

  The local $ D(\vec x)$ are expressed as a sum of a uniform part
 $\bar{D}$ and a deviations part $\Delta D(\vec{x}) $:
\begin{equation}
 D(\vec{x})=\bar{D} +\Delta D(\vec{x})=\bar{D}+2\delta D(\vec{x})
,\label{eq:coeff}
\end{equation}
\begin{equation}
 \langle \delta D(\vec{x}) \delta D(\vec{y}) \rangle=\langle \delta D^{2}
\rangle \delta (\vec{x}-\vec{y})
,\label{eq:2ndcul}
\end{equation}
  (where $\delta D(\vec{x})=\frac{\Delta D(\vec{x})}{2}$   was introducted for
 computational convenience).
   This model was considered somewhat trivial is because for
 weak disorder,
    $\frac{\Delta D}{\bar{D}}<<1$,   there is an asymptotic normal diffusive
 behavior at any dimension $ d>0$ (the irrevlevance of the disorder near the
 pure diffusive fixed-point is explicitly shown below).

   However, this does not preclude a non-trivial behavior for the strong
 disorder regime. The existence of such a regime will be manifested by a
 finite basin of attraction of the pure diffusive fixed-point in the space of
 all possible probability
distribution $P( D)$  for the locally uncorrelated diffusion constants. This
 space may be represented by the couplings $ g^{(k)}$  related to the cumulants
 of the distribution  by:
\begin{eqnarray}
{\langle \delta D(\vec{x}) \delta D(\vec{y})\rangle}_{c} &=& 2!g^{(2)}
\delta(\vec{x}=\vec{y}),\nonumber \\
{\langle \delta D(\vec{x}) \delta D(\vec{y}) \delta D(\vec{z})\rangle}_{c} &=&
3
!(-1)g^{(3)} \delta (\vec{x}=\vec{y}=\vec{z}),\nonumber \\
   \cdots & & \cdots \label{eq:cul2} \\
 {\langle \delta D (\vec{x}_{1}) \cdots \delta D(\vec{x}_{k}) \rangle}_{c} &=&
k
! (-1)^k g^{(k)} \delta (\vec{x}_1=\vec{x}_2= \cdots  =\vec{x}_k). \nonumber
\end{eqnarray}

The central question we set ourselves to address in this work is whether there
 exists a non-trivial distribution  that will be associated with the "critical"
 behavior on the separatrix boardering this basin of attraction in the
cumulants
  infinite-dimensional space. Of course there may be more than one such
 distribution or even a whole family of them. That depends on the
 renormalization-group flow on the separatrix ("critical manifold") itself.

 However, even the existence of such one scale invariant distribution is far
 from trivial since the theory may be either non-renormalizable (if an
 infinite number of new relevant couplings are generated) or may be
 renormalizable in a larger "unphysical" space (if relevant  operators,
 distinct from the above cumulants, are generated).

      Similiar questions have arisen recently in the study of quantum diffusion
(localization)
  where it was shown that the theory is renormalizable if the whole
 distribution of the conductance is considered [8-10]. The distribution takes
 three different forms depending on the system being metallic, insulating, or
 at criticality in between [8].
   Similar questions of renormalizability always arise when a random system is
 considered near its lower critical dimension. For non-random models symmetry
 ensures the full renormalizibity near the lower critical dimension
in which they have an infinite number of marginal operators mixed by the RG
flow
s,
 (e.g. $O(M)$ symmetry [11] for the
$M$-component Heisenberg model in $d=2+\epsilon$). For random  systems the
 effective replicated Lagrangian has a larger number of couplings without
 having larger symmetries to impose constraints on the renormalized couplings
 to be related to each
other  (as to preserve these symmetries). So the question of whether random
 systems are renormalizable near their lower critical dimension is much more
 delicate. While the quantum diffusion in $d=2+\epsilon$ is an example of a
 renormalizable theory [8-10]
, the random-field $O(M)$ model in $d=4+\epsilon$ provides a counterexample
 [12].

  We therefore hope that our systematic study of another such system will help
 to shed more light on this puzzling question.

   The paper is organized as follows: In the next chapter (II) the
 n-replicated Lagrangian will be derived.
 From this Lagrangian, which has the cumulants of the initial distribution
related to its bare couplings, all correlation functions may be derived. In
chapter III
we analyze the so-called annealed approximation $(n=1)$ first by the standard
 RG approach and then by a search of the universal distribution based on
 symmetry. Chapter IV is devoted to  the study of the quenched average
 $(n\rightarrow 0)$.
The problems encountered are discussed. In chapter V we summarize the
 implications of these investigations for the diffusion problem in particular
 and for random systems in general.

\vskip 15pt
\noindent{\bf \large II.  The replicated Lagrangian. }

  Without losing generality, we assume all the particles to be at
 $\vec{x}=\vec{0}$  at the initial time $t=0$, namely:
\begin{equation}
\rho(\vec{x},0)=\delta (\vec{x})
.\label{eq:density}
\end{equation}
 \ The Laplace transform of $\rho(\vec{x},t)$   is defined as:
\begin{equation}
\tilde{\rho}(\vec{x},m^{2})=\int_{0}^{\infty} dt \rho(\vec{x},t) e^{-m^{2}t}.
\label{eq:f3}
\end{equation}
$\tilde{\rho}(\vec{x},m^{2})$  obeys the equation:
\begin{equation}
m^{2}\tilde{\rho}(\vec{x},m^{2})-\rho(\vec{x},0)=\vec{\nabla} \cdot [D(\vec x
)\
vec{\nabla}]\tilde{\rho}(\vec x,m^2) ,\label{eq:diff1}
\end{equation}
or:
\begin{equation}
m^2\tilde{\rho}(\vec x,m^2)-\vec{\nabla} \cdot
[D(\vec{x})\vec{\nabla}]\tilde{\rho}(\vec{x},m^{2})=\delta(\vec{x}).\label{eq:diff2}
\end{equation}

 We may identify $\tilde{\rho}(\vec{x},m^{2})$   with the Green's function
\begin{equation}
\tilde{\rho}(\vec{x},m^{2})=\tilde{G}(\vec{x},m^{2})
,\label{eq:qr1}
\end{equation}
 which is the Laplace transform of the time domain Green's function
\begin{equation}
G(\vec{x},t)=\langle \rho (\vec{x},t) \rho (\vec{0},0)\rangle
.\label{eq:gr2}
\end{equation}

  All scaling properties may be extracted from the large $\vec x$ and $t$ (or
 small $m^2$)  behavior of these Green's functions.
 To be able to average these correlation functions over the realizations of the
 disorder we first express them as a functional integral in the standard form:

\begin{equation}
\tilde{G}(\vec{x},m^{2})=\frac{\int {\cal D}\phi \phi(\vec{x}) \phi(\vec{0})
e^{
-{\cal L}[\phi]}}{\int {\cal D}\phi e^{-{\cal L}[\phi]}},\label{eq:fu1}
\end{equation}
where:
\begin{equation}
{\cal L}[\phi]=\int d^{d}x \phi [\frac{m^{2}}{2} -
\vec{\nabla}\cdot(\frac{D(\vec{x})}{2} \vec{\nabla})]\phi,\label{eq:fu2}
\end{equation}
 or, after integration by parts and neglect of an unimportant boundary term:

\begin{equation}
{\cal L}[\phi]=\int d^dx \{\frac{m^2}{2} \phi^{2}+\frac{D(\vec{x})}{2}
(\vec{\nabla}\phi)^{2}\}.\label{eq:L3}
\end{equation}
 
The averages over the disorder may be obtained by utilizing the replica trick
 . The n-replicated partition function is obtained by including $ n$ fields
 $\phi^{\alpha}(\vec{x}) ,$ ($\alpha=1,2,....,n$):
\begin{equation}
Z^n=\int \prod_{\alpha} {\cal D} {\phi}^{\alpha}(\vec{x}) e^{- {\cal
L}[{\phi}^{
\alpha}]}
,\label{eq:trick}
\end{equation}
with:
\begin{equation}
{\cal L}^n [{\phi}^{\alpha}]=\int d^{d}x \{\sum_{\alpha} [\frac{m^{2}}{2}
({\phi
}^{\alpha})^{2} +\frac{D(\vec{x})}{2} ({\vec{\nabla}\phi}^{\alpha})^{2}]\}
,\label{eq:tr1}
\end{equation}
   Since $\lim_{n\rightarrow 0} \langle Z^{n} \rangle _{disorder}=1$, we may
 eliminate the denominator in Eq.(\ref{eq:fu1}), and calculate the average
Green's function from:
\begin{equation}
 \langle \tilde{G}(\vec x,m^2) \rangle _{disorder} =\lim_{n\rightarrow 0 }\int
\
prod_{\alpha} {\cal D} \phi^\alpha (\vec x) \phi^1(\vec x) \phi^1(\vec 0)
e^{-{\
cal L}^n[\phi^\alpha]}
.\label{eq:tr2}
\end{equation}

    Since the quenched average is usually difficult to perform the "annealed"
 approximation is often used. It consists in averaging independently the
 numerator and the denominator in Eq.(\ref{eq:fu1}).
This is also equivalent to keeping a single replica or averaging $Z^{n}$ in
 Eq.(\ref{eq:trick}) with $n=1$.
 Therefore we shall, in the rest of this chapter, analyze the generalized field
threories with ${\cal L}^{n}[{\phi}^{\alpha}]$  for any $n$.  The results of
 specific cases of annealed $(n=1)$ and quenched $(n\rightarrow 0 )$ average
 are discussed in the next two  chapters.
   Let us use the vector notation
 $\vec{\phi}=({\phi}^{1},{\phi}^{2},\cdot \cdot \cdot \cdot,{\phi}^{n})$.
 The replicated Lagrangian may be separated into a free part and an interacting
 part,  ${\cal L}^{n}={\cal L}_{o}+{\cal L}_{in} $
  where:
\begin{equation}
 {\cal L}_{0}=\frac{1}{2}\int d^{d}x \{\bar{D} (\vec{\nabla}\vec{\phi})^{2}
+m^{
2}(\vec{\phi})^{2} \}
,\label{eq:lo}
\end{equation}
and:
\begin{equation}
{\cal L}_{in}=\int d^{d}x \{\delta D(\vec{x}) (\vec{\nabla} \vec{\phi})^{2} \}
.\label{eq:lin}
\end{equation}

  The average over the disorder yields:
\begin{equation}
\langle e^{-{\cal L}_{in}} \rangle _{disorder}=e^{\int d^d x
\sum_{k=2}^{\infty}
 g^{(k)}[(\vec{\nabla}\vec{\phi})^{2}]^{k}}
,\label{eq:alin}
\end{equation}
  where $g^{(k)}$   are the couplings defined in Eq.(\ref{eq:cul2}).
  We therefore have to apply the RG analysis to the following averaged
 partition function:
\begin{equation}
\langle Z^n \rangle _{disorder}=
\int \prod_\alpha {\cal D}\phi^\alpha(\vec{x})
\exp\{-\int d^{d}x (\frac{1}{2}[m^{2}{\phi}^{2}+
\bar{D}(\vec{\nabla}\vec{\phi})^{2}]-
\sum_{k=2}^{\infty} g^{(k)}[
(\vec{\nabla}\vec{\phi})^{2}]^{k})\},\label{eq:hhh}
\end{equation}
for simplicity we shall choose $\bar D =1$ in the forthcoming calculations.
 Also in the renormalization scheme we choose to keep this term constant. To
 make it dimensionless
 $\phi$ should carry dimension of $[\phi]=L^{1-\frac{d}{2}}$ .
Hence $[(\vec{\nabla}\vec{\phi})^{2}]=L^{-d}$, and $[g^{(k)}]=L^{(k-1)d}$.

   Therefore under rescaling $L\rightarrow L/b $
, $g^{(k)}\rightarrow  b^{-(k-1)d} g^{(k)}$ and $g^{(k)}$   are irrelevant
 near the free gaussian theory for any dimension $d>0$.  We also observe that
 $ d=0$ is the critical dimension at which these couplings become marginal.
 The natural small parameter will  thus
be $\epsilon =d$.
  In the next chapters we go beyond the dimensional analysis by utilizing
 first the  RG approach and then an alternate approach based on possible
 symmetries of the non-trivial fixed-point distribution.
\vskip 15pt

\noindent {\bf \large III. The annealed (n=1) approximation.}

We begin our analysis with the consideration of the simpler annealed
 approximation. That corresponds to single component field
 $\phi$. We begin by demonstrating the problems that arise in taking the
   RG route and then show how these difficulties may be
 circumvented to find the scale-invariant distribution.

 \noindent 1. \underline{Renormalization-group approach.}

To go beyond the naive dimensional analysis we need to expand
 $e^{\{-\sum_{k} g^{(k)} [(\vec{\nabla}\phi)^{2}]^{k}\}}$
 in a power series, separate and integrate momenta
$\Lambda /b <q<\Lambda $ (where $\Lambda =1/a$ is the boundary of the Brillouin
 zone) using $e^{-H_{0}}$, reexponentiate, and rescale all momenta
 $q\rightarrow bq$  as to obtain a new effective Lagrangian. This can be done
 diagrammatically and here we only give the recursion
relations for $ k=2,3$ to order one loop ($l=lnb$):
\begin{equation}
\frac{dg^{(2)}}{dl}=-\epsilon g^{(2)}+12 g^{(3)}+56(g^{(2)})^{2}
,\label{eq:k2}
\end{equation}
\begin{equation}
\frac{dg^{(3)}}{dl}=-2\epsilon g^{(3)}
+240g^{(3)}g^{(2)}+24g^{(4)}+416(g^{(2)})
^{3}
,\label{eq:k3}
\end{equation}

 Similar expressions may be derived for all $g^{(k)}$. They will take the form:
\begin{equation}
 \frac {dg^{(k)}}{dl}=\beta _{\epsilon}^{(k)} (g^{(2)},g^{(3)}, \cdots
,g^{(k+1)
})=-(k-1) \epsilon g^{(k)}+\beta ^{(k)}_{0} (g^{(2)},g^{(3)}, \cdots ,g^{(k+1)}
),\label{eq:beta}
\end{equation}
where $\beta ^{(k)}_{0}$
 is the same beta function calculated at $\epsilon =0$. Since
 $\beta ^{(k)}_{\epsilon}$  have both positive and negative terms each of them
 has zeros. The challenge is to find at least one common zero,
 $g^{(k)}=g^{(k)\ast}$
  for all $k$,  where they all vanish simultaneously. We also note that the
  $g$'s
  will be generated under the RG recursions once the bare value of one of them
 is not zero. However, different couplings (not cumulants of $P(D)$ ) will
 not be generated and the space of all
  $ g $'s is closed under the RG transformation. These equations also tell us
 that at the fixed-point $g^{(k)\ast} \sim \epsilon ^{k-1}$, and we
may express them in terms of coefficients $a_{k}$  such that:
\begin{equation}
 g^{(k)\ast}=a_{k} \epsilon^{k-1}
.\label{eq:aa}
\end{equation}
Using the equations $\beta^{(k)}_{\epsilon} =0 $  all $a_{k}$  may, iteratively
, be related to $a_{2}$, e.g. :
\begin{eqnarray*}
a_{3} & = & \frac{a_{2}}{12}(1-56a_{2}) \\
a_{4} & = & \frac{a_{3}}{12}-10 a_{3} a_{2}-\frac{52}{3} a_{2}^{3}, \mbox{
etc
.}
\end{eqnarray*}

We can also deduce some information on the behavior away from the fixed-point.
 Suppose we try to rescale all $g^{(k)}$  multiplicatively:
\begin{equation}
 g^{(k)}=\zeta (l)^{k-1} g^{(k)\ast}
,\label{eq:gs}
\end{equation}
  very close to the fixed-point $\zeta (0) \simeq 1$, then from
Eq.(\ref{eq:gs})
:
\begin{equation}
 \frac{dg^{(k)}}{dl} = g^{(k)\ast} \frac{d[\zeta(l)^{k-1}]}{dl}= g^{(k)\ast}
(k-
1) \zeta(l)^{k-2} \frac{d\zeta}{dl}
,\label{eq:zeta}
\end{equation}
But on the other hand from Eq.(\ref{eq:beta}):
\begin{equation}
\frac{dg^{(k)}}{dl}=-\epsilon (k-1) g^{(k)\ast} \zeta ^{k-1} -\zeta^{k}
\beta_{0
}^{(k)}\{g^{\ast}\} = \zeta^{k-1}  (\zeta -1) (k-1) \epsilon g^{(k)\ast}
.\label{eq:gc}
\end{equation}
Equating the two equations (\ref{eq:zeta}) and (\ref{eq:gc})  above we obtain:
\begin{equation}
\frac{dln \zeta (l)}{dl}=(\zeta -1)\epsilon
,\label{eq: dc}
\end{equation}
from which we identify $\zeta$ as a relevant scaling field near the fixed-point
$(\zeta =1)$  with scaling exponent $\phi _{\zeta}= \epsilon$.

 \noindent 2. \underline{Invariant cumulant generating function: A search
 by symmetry.}

  Instead of looking for the invariant distribution based on the recursion
 relations (which seems pretty hopeless) we shall base our search on global
symm
etry considerations. We
shall  look first at the moment generating function:
\begin{equation}
  f(u)=\log \int_{0}^{\infty} dD P(D) e^{-Du} =\sum_{k=1}^{\infty}
(-1)^k\frac{\
langle D^k \rangle _{c}}{k!} u^k
.\label{eq:gf}
\end{equation}
Comparing with Eq.(\ref{eq:cul2}),  we have:
\begin{equation}
f(0)=0
,\label{eq:gf0}
\end{equation}
\begin{equation}
f^{'}(0)=1
,\label{eq:gf1}
\end{equation}
\begin{equation}
\frac{d^k f}{d u^k}|_{u=0}= g^{(k)} k!2^k
.\label{eq:gfk}
\end{equation}
 If we can find a function $f(u)$ such that for $2u=(\vec{\nabla}\phi)^2$ the
in
tegral:
\begin{equation}
 \tilde{{\cal L}}[f\{\frac{(\vec{\nabla}\phi)^2}{2}\}]=\int d^d x \{
f[\frac{(\vec{\nabla}\phi)^2}{2}]\}
,\label{eq:inv}
\end{equation}
   will be invariant under renormalization, then the corresponding $P(D)$  (to
 be obtained by inverse Laplace transform of $e^{f(u)}$ ) will also be
invariant
{}.

  So the problem has been reduced to finding the function $f(u)$  for which
 $\tilde{{\cal L}}(f)$ is invariant.
This search must be based on symmetry: one has to look for a
 function $f(u)$ for which the integral in Eq.(\ref{eq:inv})  is an invariant
un
der a symmetry
 operation which itself is  preserved under the RG iterations (this insight
 comes from what is known about the role of the symmetry for RG near the lower
 critical dimension [11]).
 Since $\phi $  is a scalar this cannot be  strictly an internal symmetry. It
 should be a symmetry that mixes the order parameter $\phi (\vec{x})$  and the
 coordinates $ \vec{x}$.
  The simplest one is to add  $\phi$ as the ($d+1$)th  component of a new
vector
:
\begin{equation}
 x^{\mu}=(\vec x,\phi (\vec{x})).
\label{eq:x}
\end{equation}
 The simplest invariant is then just the total arc length in this space:
\begin{equation}
 {\cal S}[f\{\frac{\vec{(\nabla}\phi)^2}{2}\}]=\int d^d x
\sqrt{1+(\vec{\nabla}\
phi)^2}
,\label{eq:inter}\
\end{equation}
which is invariant under rotations in the $d+1 $ -space, and is therefore a
natu
ral
candidate for $\tilde{{\cal L}}[f]$ in Eq.(\ref{eq:inv}).

 This type of actions has been introduced and studied by Wallace and Zia [13]
 to model
interfacial fluctuation of Ising systems in $d+1 $  dimension (in this picture
 $\phi (x)$ is the height of the interface, it carries dimensions of length,
 and the overall action is made dimensionless by
 giving  dimension of $L^{\epsilon}$  to the temperature).

  The inverse Laplace transform of $ f(u) \sim e^{[-(1+2u)^{1/2}]}$   is
 the inverse gaussian distribution [14].
  This distribution has two free parameters which may be related to its
 average and its variance, in terms of which the universal renormalized
 distribution is :
\begin{equation}
 P_{u} (D)=(\frac{1}{2 \pi \langle \Delta D^{2} \rangle})^{1/2}
(\frac{\bar{D}}{
D})^{3/2} \exp\{ - \frac{\bar{D}}{2 \langle \Delta D
^{2} \rangle} \frac{(D-\bar{D})^{2}}{D} \}
.\label{eq:dd4}
\end{equation}
  The form of this distribution remains invariant under the RG flow. The only
 parameter which changes is the ratio:
\begin{equation}
R=\frac{\langle \Delta D^2 \rangle}{\bar{D}^2}=\frac{8g^{(2)}}{\bar{D}^2}
.\label{eq:dg}
\end{equation}
Higher cumulants are expressed in terms of $R$ and $\bar{D}$ :
\begin{equation}
\langle \Delta D^{n} \rangle _{c}=(2n-3)!! R^{n-1} \bar{D}^{n}
.\label{eq:rs}
\end{equation}
 The average $\bar{D}$ is a redundant parameter which can be given any value
 $\bar{D} > 0$.

 On the renormalized trajectory  the RG flow for $R$  are determined by the
 equation:
\begin{equation}
\frac{dR}{dl}=-\epsilon R + R^2
,\label{eq:rr}
\end{equation}
which has $R^{\ast}=\epsilon$  as its fixed-point.
  Thus the fixed-point distribution is (choosing $\bar{D}=1$ ):
\begin{equation}
 P_{\epsilon}^{*} (D)=(\frac{1}{2 \pi D^3 \epsilon})^{1/2}
\exp\{-\frac{1}{2\epsilon} \frac{(D-1)^2}{D} \}
,\label{eq:fp}
\end{equation}
this distribution with $\epsilon =1 $ is plotted
in Fig.1.

For $R< \epsilon $, the RG flow will be to smaller $R$ or
 $<\Delta D^2> \rightarrow 0 $ while keeping $\bar{D}$  fixed. The flow takes
th
e distribution to:
\begin{equation}
 \lim_{R\rightarrow 0}(\frac{\bar{D}}{2 \pi D^3 R})^{1/2} \exp{\{-\frac{1}{2R}
\
frac{(D-\bar{D})^2}{D \bar{D}} \}}=\delta (D-\bar{D})
,\label{eq:lim}
\end{equation}
as expected.

For $R> \epsilon $ , the RG flow will be towards larger $R$  .
 The moments $< \Delta D^k > $  diverage as $R^{k-1}$  and the diffusion
 process is ill-defined. (As $R \rightarrow \infty,$ $P(D) \sim D^{-3/2}$ for
$R^{-1}<<D <<R$).

   It is also straightforward to identify
 $\zeta (l) $ disscussed above as the scaling field $\zeta (l)=R(l)/R^{\ast}$.
 Hence the crossover exponent for $\Delta R \sim R-R^{\ast}$  is $\epsilon$
 . This is the only relevant direction near the fixed-point
 [15] and any variation from the fixed-point along other directions  is
 irrelevant,
  namely the fixed-point is stable in all other directions in this parameter
 space and the critical manifold (the separatrix) has codimension one.
Operators
 which break the
 $\phi \rightarrow  - \phi$
  symmetry or which contain power of $\phi$, rather than of its gradient, are
 also relevant [15] but  are not important
in the present context (besides $m^2 \phi ^2$ which is discussed next).

\noindent 3.\underline{Anomalous diffusion on the critical manifold.}

For any distribution on the critical manifold the asymptotic scaling behavior
 will be determined  by the RG flows near the fixed-point.
These are expected to be different than the simple diffusion which occurs near
the free gaussian fixed-point.

At the fixed-point the field $\phi^{2}$
  will acquire anomalous dimension in order to keep $\bar{D}$ (and all the
 rest of the couplings) fixed. The anomalous dimension will be exactly  that
 associated with scaling field $\zeta^{-1} (l)$  at the fixed-point may be
 interpreted as the renormalization factor which multiplies $\phi^{2}$ to keep
 the couplings fixed. Since $\zeta (l) \sim e^{\epsilon l} =b^{\epsilon}$, the
 rescaling of  $\phi ^2$ will be :
\begin{equation}
 \phi^{2} \rightarrow \zeta ^{-1} \phi^2 \sim b^{d-2-\epsilon } \phi^2
.\label{eq:sdc}
\end{equation}
That induces an anomalous rescaling of the Laplace transform parameter $m^2$ :
\begin{equation}
m^2 \rightarrow m^2 b^{2+\epsilon}
,\label{eq:fgr}
\end{equation}
which by its definition scales as $1/t$.
 Hence  the time $t$ will rescale as:
\begin{equation}
t\rightarrow t b^{-2-\epsilon}
,\label{eq:ssd}
\end{equation}
while length rescales as $b^{-1}$, and therefore:
\begin{equation}
\langle \vec{x}^2 \rangle \sim b^{-2} \sim t^{\frac{2}{2+\epsilon}}
,\label{eq:chy}
\end{equation}
Hence the diffusion is anomalous with exponent $\mu$ in Eq.(\ref{eq:asym})
given
 by:
\begin{equation}
2\mu=\frac{1}{1+\frac{\epsilon}{2}} \sim 1- \frac{\epsilon}{2}
.\label{eq:fty}
\end{equation}

Although locally stable we cannot prove that this fixed-point is unique.
This question could be explored looking for all possible functions $f((\nabla
\phi)^2)$ which are scale invariant. One way to approach the problem is to
ask what are the non-linear symmetries preserved under RG. The "Euclidean"
 symmetry is one of them and the action in Eq.(36) is its only invariant.
 We cannot rule out, however, the possibility of other preserved symmetries
with
 other invariants. It will be very interesting if this general question could
 be investigated more systematically.  If other fixed-points exist they may
 not follow the simple gap scaling above, and a "multifractal" behavior
 cannot be ruled out.
\vskip 15pt

 \noindent {\bf \large IV.  The quenched average $ (n \rightarrow 0)$. }

  Our goal is to pursue the same succesful route we followed to the solution
 in the annealed approximation, for the quenched case as well.
       We therefore repeat the RG calculation for general $ n $ $ (n\neq 1)$
 to be analytically continued to zero. Here, however, we encouter a new
 difficulty: The space spanned by the couplings  $g^{(k)}$ is not closed under
 renormalization.

    We shall explain it for terms to order $(\partial \phi )^{4}$ but similar
 behavior  takes place for higher powers as well (see Appendix).
 The term in the bare Lagrangian is of the form:
\begin{displaymath}
 g^{(2,0)}(\sum_{\alpha =1}^{n} \sum _{i=1}^{d} \partial_{i} \phi^{\alpha}
 \partial_{i} \phi^{\alpha})^2=g^{(2,0)} \sum_{\alpha,\beta =1}^{n}
\sum_{i,j=1}
^{d} \partial_{i} \phi^{\alpha} \partial_{i}
\phi ^{\alpha} \partial_{j} \phi ^{\beta} \partial _{j} \phi^{\beta}
{}.
\end{displaymath}
 To order $ (g^{(2,0)})^2$  the contraction of two such terms give rise to a
 renormalization of $g^{(2,0)}$  itself but also to the generation of term of
 the form:
\begin{equation}
g^{(0,2)} \sum_{\alpha,\beta} \sum_{i,j} (\partial_{i} \phi^{\alpha}
 \partial_{i} \phi^{\beta})(\partial_{j} \phi^{\alpha} \partial_{j}
\phi^{\beta}
)
,\label{eq:phi}
\end{equation}
The complete recursion relations are given in the Appendix.
Clearly such a term takes us outside of the realm of the diffusion model we
 began with.
   If we insist on still following the RG flow in the a larger number of
unphys
ical
 parameter space we can generalize the approach based
 on symmetry discussed above for the case $n=1$  to general $n$, by looking
 at a vector in the
 $d+n$  dimension space
 $(x_1,\cdots ,x_d, \phi^1 (\vec x), \cdots ,\phi ^n (\vec x) )$.  The
 invariant area in this space is:
\begin{equation}
 {\cal S}[\phi^{\alpha}(\vec x)]=\int d^d x (det g)^{1/2}
,\label{eq:man}
\end{equation}
with
 $ g_{ij}=\delta_{ij}+\sum_{\alpha=1}^{n} \partial_{i} \phi^{\alpha}
\partial_{j
} \phi^{\alpha}. $

Although no distribution may correspond to this function (since it cannot be
 a cumulant generating function) there is a non-trivial fixed-point for finite
 $n$.  The incomplete
 one-loop analysis of Lowe and  Wallace [16] yields
 $R^{\ast} = \epsilon /n$ and will diverges in the $ n\rightarrow 0$ limit. If
 this behavior will surive for a larger numbers of loops, it  would imply that
e
ven in the
 larger space the trivial fixed-point is the only one accessible by the
  RG approach
(which still leaves the possibilty of a non-perturbative strong coupling
 behavior).
\vskip 15pt

\noindent {\bf \large  V. Conclusions.}

  In view of the new understanding acquired in the quantum diffusion problem
 on the importance of renormalizing the full conductance distribution [8-10]
 (rather than the first two moments alone), we have addressed the question of
 the distribution of   the  diffusion
constant in  random classical diffusion. We have chosen here the simplest
 model for diffusion in disordered media which has $d=0$  as its critical
dimension. Our analysis was performed on the replicated Lagrangian in which
 the cumulants of the distribution are related to the coupling constants.

  We first looked at the annealed approximation  for which we could identify
 the invariant distribution. We have found a non-trivial fixed-point along
 the renormalized trajectory separating a free gaussian diffusion fixed-point
 for small $\frac{<\Delta D^2>}{\bar{D}^2}$ from a regime
where the cumulants ratio divgerges
 $<\Delta D^{(k+1)}>/<\Delta D^{(k)}> \rightarrow \infty$. At the fixed-point
 itself  an anomalous diffusion $<\vec x ^2> \sim t^{1-\frac{\epsilon}{2}}$
  was found.

   For the quenched average
 $n \rightarrow 0$   our results are so far negative: the theory is not
 renormalizable because unphysical terms are generated under renormalization.
 Such a behavior may occur in other random systems near their lower critical
 dimension. That may also
indicate that an even more general approach (for example including the
 possibilty of replica symmetry breaking) may be necessary.
 Other investigations of the distributions of the appropriate physical
 quantities in other random models may shed more light and will be most
 worthy.
\vskip 15pt
\newpage
{\bf Acknowledgments}

 We are very thankful to R.K.P. Zia for most useful discussions. This work was
 partly supported by a grant from the Corporate Research Laboratories of the
 Eastman-Kodak Company.
\newpage
{\bf Appendix}

In this appendix, we present the recursion relations of two lowest order
 couplings in  Eq.(\ref{eq:man}) for general $n$ approach. As mentioned in
chapter IV, we need to consider other  (cross) terms like
the one in Eq.(\ref{eq:phi})  which does not appear in the
original Lagrangian in Eq.(\ref{eq:hhh}). Including the terms generated, one
 should rewrite the Lagrangian as follows:
\begin{eqnarray}
{\cal L}[\phi^{\alpha}(\vec{x})]&=&\int d^dx \frac{1}{2} \sum_{\alpha,i}
\partial_{i}\phi^{\alpha} \partial_{i}\phi^{\alpha} -\sum_{\alpha,\beta} \sum_{i,j}[
g^
{(2,0)}(\partial_{i}\phi^{\alpha} \partial_{i} \phi^{\alpha} \partial_{j}
\phi^{
\beta}
\partial_{j} \phi^{\beta})
+g^{(0,2)} (\partial_{i} \phi^{\alpha} \partial_{i} \phi^{\beta}
\partial_{j}\phi^{\alpha} \partial_{j} \phi^{\beta} )]  \nonumber \\
&&
-\sum_{\alpha,\beta,\gamma} \sum_{i,j,k} [ g^{(3,0)}(\partial_{i}\phi^{\alpha}
\partial_{i} \phi^{\alpha})(\partial_{j}\phi^{\beta}
\partial_{j}\phi^{\beta})(\
partial_{k}\phi^{\gamma}\partial_{k}\phi^{\gamma})
 \nonumber \\ & &+g^{(1,2)}(\partial_{i}\phi^{\alpha} \partial_{i}
\phi^{\alpha}
)(\partial_{j} \phi^{\beta} \partial_{j}\phi^{\gamma})
(\partial_{k}\phi^{\beta} \partial_{k}\phi^{\gamma}) \nonumber \\  & &
+g^{(0,3)}(\partial_{i}\phi^{\alpha}\partial_{i}\phi^{\beta})(\partial_{j}\phi^{
\gamma}\partial_{j} \phi^{\alpha})(\partial_{k}\phi^{\beta}\partial_{k}
\phi^{\gamma})]+\cdots  \mbox{higher power terms}.
\label{eq:gen}
\end{eqnarray}
By  diagrammatic calculations, we found the following recursion relations for
 $g^{(2,0)}$ and $g^{(0,2)}$ ( to order one loop):
\begin{eqnarray}
\frac{dg^{(2,0)}}{dl}&=&-\epsilon g^{(2,0)} +12 g^{(3,0)}+(2n+2) g^{(1,2)}
+3g^{
(0,3)} +40(g^{(2,0)})^2 + (16n+32)g^{(0,2)} g^{(2,0)} \nonumber \\
& & +(2n+14) (g^{(0,2)})^2
,\label{eq:gen1}
\end{eqnarray}
\begin{equation}
\frac{dg^{(0,2)}}{dl}=-\epsilon g^{(0,2)} +8g^{(1,2)} +(3n+6)g^{(0,3)}
+(12n+28)
(g^{(0,2)})^2 +64g^{(0,2)}g^{(2,0)}+16(g^{(2,0)})^2
.\label{eq:gen2}
\end{equation}
 For the case $n=1$, the distinction between  the two terms in the
Eq.(\ref{eq:g
en}) with coefficients
 $g^{(2,0)},g^{(0,2)}$ disappears, so one can identify  $g^{(2)}$ in
Eq.(\ref{eq
:k2})  as the sum of $g^{(2,0)}$ and $g^{(0,2)}$ in Eq.(\ref{eq:gen}). For the
s
ame reason, $g^{(3)} $ in Eq.(\ref{eq:k2})  is the sum of $g^{(3,0)}$,
$g^{(1,2)}$ and $g^{(0,3)}$. Therefore the sum of the two equations above is
ide
ntical to Eq.(\ref{eq:k2}), for $n=1$.

\newpage
\noindent{\large\bf Figure Captions}

\noindent{\bf Fig.1}
-The form of the invariant inverse gaussian distribution (Eq.(\ref{eq:dd4}))
wit
h parameters $\bar{D}=
1$ and $<\Delta D^2>=1$ (it is the critical distribution for $d=1$ within the
an
nealed
approximation).
\vskip .5truein

\end{document}